\documentclass[12pt]{article}
\usepackage{bm}
\usepackage{cite}
\usepackage{mathrsfs}
\usepackage{slashed}
\usepackage{graphicx}
\usepackage{hyperref}
\usepackage{verbatim}
\usepackage{color}
\usepackage{enumerate}
\usepackage{authblk}
\usepackage[hang,flushmargin]{footmisc}
\usepackage{lipsum}
\usepackage[font=footnotesize]{caption}
\usepackage{soul}
\usepackage[utf8]{} 
\usepackage{empheq}
\usepackage{amssymb,amsmath,amsfonts}
\usepackage[makeroom]{cancel}
\usepackage[normalem]{ulem}
\usepackage{hyperref}
\usepackage{color}

\numberwithin{equation}{section}

\definecolor{blue-violet}{rgb}{0.54, 0.17, 0.89}
\definecolor{PineGreen}{cmyk}{0.92, 0, 0.59, 0.25}
\definecolor{OliveGreen}{cmyk}{0.64, 0, 0.95, 0.40}
\definecolor{RawSienna}{cmyk}{0, 0.72, 1, 0.45}
\definecolor{Gray}{cmyk}{0, 0, 0, 0.50}
\definecolor{MidnightBlue}{cmyk}{0.98, 0.13, 0, 0.43}
\definecolor{Orange}{cmyk}{0, 0.61, 0.87, 0}
\definecolor{LimeGreen}{cmyk}{0.50, 0, 1, 0}
\definecolor{Green}{cmyk}{1, 0, 1, 0}

\renewcommand{\tilde}{\widetilde}
\renewcommand{\hat}{\widehat}

\usepackage{mathrsfs}

\def\cJ{\mathcal{J}}

\def\cL{\mathcal{L}}

\def\cO{\mathcal{O}}

\def\be{\begin{eqnarray}}
\def\ee{\end{eqnarray}}
\def\beann{\begin{eqnarray*}}
\def\eeann{\end{eqnarray*}}
\def\beq{\begin{equation}}
\def\eeq{\end{equation}}
\def\ba{\begin{array}}
\def\ea{\end{array}}
\def\ben{\begin{enumerate}}
\def\een{\end{enumerate}}
\def\bea{\begin{eqnarray}}
\def\eea{\end{eqnarray}}

\begin{document}

\title{\vspace{-70pt} \Large{\sc Classical integrability in 2D and asymptotic symmetries}\vspace{10pt}}
\author[a]{\normalsize{Marcela C\'ardenas}\footnote{\href{mailto:marcela.cardenasl@uss.com}{marcela.cardenasl@uss.com}}}

\affil[a]{\footnotesize\textit{Universidad San Sebasti\'an, Facultad de Ingenier\'ia,  Bellavista 7, Recoleta, Santiago, Chile.}}

\date{}

\maketitle

\begin{abstract}
These lecture notes are a contribution to the proceedings of the school ``Geometric, Algebraic and Topological Methods for Quantum Field Theory",  held in Villa de Leyva, Colombia, from 31st of July to 9th of August 2023. Its intention is to put together several basic tools of classical integrability and contrast them with those available in the formulation of asymptotic symmetries and the definition of canonical charges in gauge theories. We consider as a working example the Chern-Simons theory in 3D dimensions, motivated by its various applications in condensed matter physics, gravity, and black hole physics. 

We review basic aspects of the canonical formulation, symplectic geometry, Liouville integrability, and Lax Pairs. We define the Hamiltonian formulation of the Chern-Simons action and the canonical generators of the gauge symmetries, which are surface integrals that subject to non-trivial boundary conditions, realize transformations that do change the physical state, namely large (or improper gauge transformations). We propose asymptotic conditions that realize an infinite set of abelian conserved charges associated with integral models. We review two different cases: the Korteweg-de Vries equation for its connection with the Virasoro algebra and fluid dynamics, and the Ablowitz–Kaup–Newell–Segur (AKNS) hierarchy, as it embeds an infinite class of non-linear notable integrable evolution equations.  We propose a concrete example for gravity in 3D with $\Lambda<0$, where we find a near-horizon asymptotic dynamics. We finalize offering some insights on the initial value problem, its connection with integrable systems and flat connections. We study some properties of the Monodromy matrix and recover the infinite KdV charges from the trace invariants extracted from the Monodromy evolution equation that can be written in a Lax form.

%this system of equations is  
%Concerning that the theory is trivially solved by flat connections, we connect this Considering the relation with Chern-Simons and gravity in three dimension, we construct . Meaning an infinite set of  in Gaussian normal coordinates, that has been interpreted as 

%Chern-Simons has been a qualitative model to show how asymptotic symmetries work. As opposed of sigma models, here integrable models appear imposing asymptotic onditions on the fields. Field theories of this type, having no local degrees of freedom but still globally non-trivial, are called topological field theories.

\end{abstract}

\newpage
\tableofcontents

\section{Introduction}

%These lecture notes attempt to bring together introductory tools for the study of classical integrability of 2D theories, and  bring a concrete application to 1+1 integrable systems with relevant properties and motives them from the notion of asymptotic symmetries.

These lecture notes aims to bring together introductory notions on classical integrability in 2D and asymptotic symmetries in 2+1 gauge theories. They expect to put these two lines of research in a common framework, and contribute to other active topics in physics that use elements of integrability, such as Sigma models \cite{Integrability-Sigma-Models}, String theory \cite{adscft,Integrable-Strings}, or self-dual Yang-Mills theories through the Ward's conjecture \cite{Ward:1985gz,sfyg,mw}. Here, we show how integrable models in 2D also naturally appear in Chern-Simons theories in 3D by means of its asymptotic dynamics and improper (or large) gauge symmetries. These are a class of symmetry transformations that do change a physical state and encompass a large set of inequivalent solutions sharing the same phase space. In the context of a Chern-Simons theory, the role of asymptotic degrees of freedom becomes highly relevant, as their field equations are solved by flat connections implying trivial bulk dynamics. Then, it is by means of its edge states that propagating modes manifest when manifolds are endowed with boundaries. Also, through the holonomies, that captures topological properties. 
We also expect to put side by side two viewpoints of computing observables. On the one hand, the existence of gauge symmetries generates conserved quantities based on Noether theorems. On the other, using tools of integrability and the associated monodromy matrix, they can be constructed from trace invariants expanded in the spectral parameter. Here we compute it explicitly for the Korteweg-de Vries charges.

Chern-Simons theories has prompted advances in Gravity \cite{Witten:1988hc}, Supergravity \cite{Achucarro:1987vz,Hassaine:2016amq}, quantum gravity \cite{Carlip:2004ba}, black hole physics \cite{Banados:1998gg,Riegler:2017fqv}, and condense matter physics via e.g., Hall effects \cite{Zee:1995avy,Wen:1995qn,suskind,Polychronakos,Witten:2015aoa}. Along these lecture notes, we deepen into this model because of its versatility and for the possibilities that broaden when it is seen through the lens of integrability. 
 We now give an incomplete list of articles, lectures, and books that can help to dive into these broad topics. For an introductory view on classical integrability, we recommend the books \cite{Babelon:2003,Backlund,Dickey} and the lecture notes \cite{Torrielli:2016ufi,Driezen:2021cpd}. To deepen in Canonical quantization we recommend \cite{Dirac,Henneaux:1992ig} and for general review on Noether’s theorems, gauge symmetries and boundary terms, the article \cite{Reyes}.

 %The idea of this notes is mainly to present a unifying perspective for understanding the notion of asymptotic symmetries, integrable systems and minimal elements of the scattering problem, through the properties of the associated monodromy matrix. \mnote{Hablar de gravedad}

%  Here we present a broad perspective, from with both topic can be beneffited from. 

%As a model of lower dimensional dynamics, expresses in clear manner how boundary degrees of freedom  have a rich asymptotic structure.

We give now the distribution of the manuscript. In Section \ref{Section 1}, we present introductory elements of Hamiltonian evolution and Symplectic geometry. Then, we provide a classical definition of integrability by Liouville and prove Liouville's theorem that ensures the existence of a canonical transformation where solutions of Hamilton's equations are linear in time. The notion of Lax pairs and trace invariants are also presented, which offers a field theory set up to define integrability. In Section \ref{Section 2}, we define the Chern-Simons action and its field equations. We focus on the canonical realization of its gauge symmetries and classify them according to the value of the symmetry generators. We define improper gauge transformations as the physical transformations that generate the asymptotic symmetry algebra. In Section \ref{Section 3}, we give explicit asymptotic conditions at spatial infinity for the Chern-Simons vector field and its symmetry generators and use the definition of canonical charges to realize the KdV infinite tower of conserved quantities. As a second example, we propose a larger integrable hierarchy, the AKNS non-linear equations. We discuss its Bi-Hamiltonian structure, which serves to prove the abelian commutation relation of its charges. Finally, in Section \ref{Section Flat}, we comment on the relation among flat connections, the first order problem, and Wilson lines, which are fundamental solutions to the linear problem. We study the evolution of the Monodromy matrix that can be written in a Lax form. We construct conserved quantities from its trace.

\section{Classical integrability and symplectic geometry} \label{Section 1}
We introduce basic aspects of the Hamiltonian evolution and the geometric properties of the phase space through the
Symplectic geometry. This serves to present the Liouville theorem which provides a definition of classical integrability.  The notion of Lax matrix is also discussed. The isospectral time evolution of the Lax operators and its characteristic dynamical structure are central aspects towards a field theory definition of integrability.

\subsection{Hamiltonian classical mechanics}
In classical mechanics, a dynamical system is represented by a set of points $\left(q_{i}(t),p_{i}(t)\right)$, called the coordinates and the conjugate momenta, respectively. 
They are labeled by $i=\left\{ 1,n\right\}$ where $n$ indicates the number of degrees of freedom of the system, as it is finite-dimensional. The canonical pair $\left(q,p\right)$ live in the \emph{phase space}, that is a differential manifold $M$.

A phase space is said to be \textit{symplectic},
when it is endowed with a non-degenerate Poisson structure $\left\{\,,\,\right\} $. To give the following definitions, we work on canonical coordinates, that are characterized by the fundamental operation $\left\{ p_i,q_{i}\right\}=\delta_{ij}$, which is the only non-vanishing bracket among them.

The Poisson bracket in canonical coordinates is defined as 

\begin{equation} \label{Classical-Poisson}
\left\{ f,g\right\} =\sum_{i=1}^{n}\frac{\partial f}{\partial p_{i}}\frac{\partial g}{\partial q_{i}}-\frac{\partial f}{\partial q_{i}}\frac{\partial g}{\partial{}p_{i}},
\end{equation}
where $f,g$ are two functions in $M$, whose space is defined by $\mathcal{F}\left(M\right)$. This product is an operation that satisfies anti-commutativity,
the Leibniz rule, and the Jacobi identity. The Poisson bracket \eqref{Classical-Poisson} also defines a Lie algebra
on $\mathcal{F}\left(M\right)$. 

%Having a function $f\in\mathcal{\mathcal{F}}\left(M\right),$
%there is an associated vector $X_{f}$ that can be constructed 
%\begin{equation}
%X_{F}=\left\{ F,\cdot\right\} =\sum_{i=1}^{n}\frac{\partial F}{\partial p_{i}}\frac{\partial}{\partial q_{i}}-\frac{\partial F}{\partial q_{i}}\frac{\partial}{\partial q{}_{i}}.
%\end{equation}

%The map $F\mapsto X_{F}$, defines an homomorphims of Lie algebras\footnote{An homomorphims of Lie algebras is a morphism (an application or a
%map) that preserves the algebraic structure } $\left[X_{F},X_{G}\right]=X_{\left\{ F,G\right\} }$.
The evolution of functions on $\mathcal{F}\left(M\right)$
is generated by another function of the phase space called Hamiltonian
$H(q_{i},p_{i})$,
\begin{equation}
\frac{\partial q_{i}}{\partial t}=\frac{\partial H}{\partial p_{i}}\qquad,\qquad\frac{\partial p_{i}}{\partial t}=-\frac{\partial H}{\partial q_{i}}.
\end{equation}

For any function $F\in\mathcal{F}\left(M\right)$, the above first-order equations, namely, the Hamiltonian equations, imply
\begin{equation}
\frac{\partial F}{\partial t}=\left\{ H,F\right\} .
\end{equation}

 One can infer that $H(q,p)$ is automatically conserved under time evolution. Every $F$ that commutes with $H$ is said to be a conserved quantity. The trajectories that make $H$ (or F) a constant a constant in time, foliate a submanifold of the phase space.

%In the case of $X_{H}\left(F\right)=0$, $F\left(state\right)$ is
%a constant  in time $f$, meaning that $F$ is a conserved quantity.
%The trajectories that make $F$ a constant, foliate a submanifold of %the phase space $M_{f}$. 

\subsection{Symplectic geometry}
The existence of a symplectic manifold $M$ is in one-to-one correspondence with the definition of differential 2-form $\omega$.  In canonical coordinates, the symplectic two-form is given by,
\begin{equation}\label{canonical-symplectic}
\omega=\sum_{i=1}^{n}dp_{i}\wedge dq_{i},
\end{equation}
where the exterior derivative $d$ is operating on the associated phase space. It is immediate to show that $\omega$ is closed, i.e. $d\omega=0$, and that one can define the canonical one-form, 
\begin{equation} \label{presimplectic}
\alpha=\sum_{i=1}^{n}p_{i}dq_{i},
\end{equation}
such that $\omega=d\alpha$.

We define the vector $X_{f}$,
\begin{equation}\label{Hamiltonian vector}
X_{f}=\left\{ f,\cdot\right\} =\sum_{i=1}^{n}\frac{\partial f}{\partial p_{i}}\frac{\partial}{\partial q_{i}}-\frac{\partial f}{\partial q_{i}}\frac{\partial}{\partial q{}_{i}}.
\end{equation}
The map $f\mapsto X_{f}$, defines an homomorphism of Lie algebras\footnote{An homomorphism of Lie algebras is a morphism (an application or a
map) that preserves the algebraic structure. } $\left\{X_{f},X_{g}\right\}=X_{\left\{ f,g\right\} }$.
It is possible to show that 
\begin{equation}\label{symplectic-from-vector}
  \left\{ f,g\right\} =\omega\left(X_{f},X_{g}\right)  
\end{equation}
This property immediately leads to $\omega\left(X_{f},X_{g}\right)=-\omega\left(X_{g},X_{f}\right).$
The symplectic form is by construction, invariant under coordinate transformations and is therefore globally well-defined.

The invariant action of a vector $X_{f}$ over $\omega$ imply the condition , 
\begin{equation} \label{derivadalie}
   \mathcal{L}_{X_{f}}\omega= (d\,\iota_{X_{f}}+\iota_{X_{f}}\,d)\omega\equiv0.
\end{equation}
where $ \mathcal{L}_{X_{f}}= (d\,\iota_{X_{f}}+\iota_{X_{f}}\,d)$ is the Lie derivative and $\iota_{X_{H}}$ is the interior product. Since $\omega$ is exact, then condition of invariance on $\omega$ \eqref{derivadalie} reduce to $d(\,\iota_{X_{f}}\omega)=0$, or equivalently, to
\begin{equation}\label{invariancecondition}
    \iota_{X_{f}}\omega=-dH.
\end{equation}
The previous equation establishes that finding a Hamiltonian function $H$ is equivalent to proving the invariance of $\omega$. When this is the case, it is said that the action of $X_{f}$ on the phase space $M$ is symplectic. The vector $X_{H}$ is called the \textit{Hamiltonian flow} an immediately satisfies \eqref{derivadalie}.

%When a system is endowed with a number of conserved quantities, the phase space acquires a structure, such that the motion is constrained to certain orbits.

\subsection{Liouville Integrability}

A classical dynamical system is said to be integrable in the Liouville
sense if there are $n$ phase space functions $F_{i}$ , such that
\begin{equation}
\frac{\partial F_{i}}{\partial t}=\left\{ H,F_{i}\right\} =0\quad\text{and}\quad\left\{ F_{i},F_{j}\right\} =0,
\end{equation}
for all $i,j\in\left\{ 1,\ldots,n\right\} $.
%The property of involution can be showed (maybe) from the Jacobi identity???

In addition, these conserved quantities are linear independent among
themselves, which is 
a consequence of the linear independence of the one-forms
\begin{equation}
dF_{i}=\frac{\partial F_{i}}{\partial q_{j}}dq^{j}+\frac{\partial F_{i}}{\partial p_{j}}dp^{j}.
\end{equation}
The independence means that at generic points (i.e. anywhere except
on a set of measure zero), the $dF_{i}$ are linearly independent,
or that the tangent space of the surface $F_{i}$ exists everywhere
and has dimension $n$. \emph{There cannot be more than $n$ independent
quantities in involution; otherwise, the Poisson bracket would be degenerate.}
The last statement also implies that the Hamiltonian $H$ is a function of the $F_{i}$, as it is also a conserved quantity. 

 \subsubsection*{Liouville theorem} The solution of the equations of motion of a Liouville integrable system can be obtained by the construction of \emph{quadratures}, which are generating functions of a particular type of canonical transformation, 
\begin{equation} \label{generating}
S(F,q)=\int_{M_{f}}\alpha=\int_{q_{0}}^{q} p_{i}\left(F_{j},q_{j}\right)dp'_{i},
\end{equation}
where $\alpha$ is the symplectic 1-form \eqref{presimplectic}. The integral is evaluated on two given points of the submanifold $M_{f}$ of the phase space in which one can solve the momenta as $p_{i}=p_{i}(F_{j},q)$.
The Liouville theorem shows that $S$ is a well-defined function, and it does not depend on the path. Using Stokes’ theorem, this comes as a consequence of showing that $d\alpha=0$ on $M_{f}$.

\emph{Proof:}
We consider the transformation
$$\left(p_{i},q_{i}\right)\longrightarrow\left(F_{i},\psi_{i}\right),$$ where $F_{i}$ are conserved quantities.
We take the vector $X_{F_{i}}$, following the definition in \eqref{Hamiltonian vector}, and consider the operation,
\begin{equation}
   X_{F_{i}}F_{j}= \left\{ F_{i},F_{j}\right\}=0.
\end{equation}
The above implies that the vectors $X_{F_{i}}$ are tangent to the manifold $M_{f}$ and since $F_{i}$ are independent among themselves, they form a basis of $M_{f}$. Using them as a basis, one construct the symplectic two-form $\omega\left(X_{f},X_{g}\right)$ that provided \eqref{symplectic-from-vector}, vanishes.

We consider now
\begin{equation} 
\delta S=S(\gamma')-S(\gamma)=\int_{\gamma'}\alpha-\int_{\gamma}\alpha
\end{equation}
which is the difference of the quadrature, evaluated in two curves $\gamma'$ $\gamma$, where both share the same initial and ending points on $M_{f}$. Then, the integral can be seen as performed in a closed path $\partial\Sigma$, such that infinitesimally 
\begin{equation}
d S=\oint_{\partial \Sigma}d\alpha=\int_{\Sigma}\omega=0
\end{equation}
where $\Sigma$ is the surface enclosed by $\partial\Sigma$. Then, considering $\omega=0$ evaluated on $M_{f}$, we have that $S$ exists and is (locally) well-defined.
As a result, the differential

 \begin{equation}
d S = \psi_{i}dF_{i}+p_{i}dq_{i},\quad\text{but}\quad d^{2} S =0,\quad\text{then}\quad dp_{i}\wedge dq_{i}=dF_{i}\wedge d\psi_{i} .
\end{equation}
We consider now the symplectic two-form, that transforms preserving its structure,
\begin{equation}
\omega=\sum_{i=1}^{n}dp_{i}\wedge dq_{i}=\sum_{i=1}^{n}dF_{i}\wedge d\psi_{i}.
\end{equation}
Then, the transformation is canonical and the Hamiltonian equations in these new coordinates are trivial
\begin{equation}
\frac{\partial F_{i}}{\partial t}=\left\{H,F_{i}\right\}=0\qquad,\qquad\frac{\partial \psi_{i}}{\partial t}=\left\{ H, \psi_{i}\right\}=\frac{\partial \psi_{i}}{\partial F_{i} }=\Psi_{i},
\end{equation}
as $\Psi_{i}$ are only functions of $F_{i}$, so they are constants in time as well.

%If \eqref{generating} does not depend on the path, the the function $\psi_{i}=\frac{\partial S}{\partial p_{i}}$ exists.
 
 % The problem now resides in finding such canonical transformation. For doing so, one construct a generating function $S$. The first requirement is to assume there is a submanifold $M_{f}$ of the phase space defined by $F_{i}=f_{i}$, in which one can solve the momenta as $p_{i}=p_{i}(f,q)$. Then, the generating function

 %\begin{equation} \label{diferential S}
%S(F,q)=\int\alpha=\left.\int p_{i}\left(F_{j},q_{j}\right)\right|_{F}dp_{i},
%\end{equation}

 %The proof of their existence is fully derived in \cite{Babelon:2003}, while here we present the main result. Quadratures ensure being able to find the canonical transformation
%$$\left(p_{i},q_{i}\right)\longrightarrow\left(F_{i},\psi_{i}\right),$$ where $F_{i}$ are conserved quantities.
%We consider now the symplectic two-form, that transforms preserving its structure,
 %\begin{equation}
%\omega=\sum_{i=1}^{n}dp_{i}\wedge dq_{i}=\sum_{i=1}^{n}dF_{i}\wedge d\psi_{i}
%\end{equation}

\emph{In this basis the equations of motion are decoupled. The solutions are trivially found because $F_{i}$ 
are constants and $\psi_{i}$ are linear functions in time. } 
\subsection{Lax Pairs and a field theory notion of integrability} \label{subsection: Lax Pairs}
Let us consider two matrices, $M$ and $L$, that depend on functions living in a phase space, i.e., in $\mathcal{F}\left(M\right)$. They are related by the equation
\begin{equation}\label{laxeq}
    \dot{L}=[M,L].
\end{equation}
Here $[\,,\,]$ denotes the commutator of matrices. Recasting the equations of motion of a Hamiltonian system in the above form, namely, as a Lax Pair, immediately implies the existence of infinite conserved charges. Let us define $H_{n}=\text{Tr}(L^{n})$ where $n\in \mathbb{N}$, and perform the operation
\begin{equation}
   \dot{L^{n}}=[L^{n},M]=\sum_{j=0}^{n-1}L^{j}[M,L]L^{n-j-1}.
\end{equation}
By tracing the above equation, expanding the sum, using the cyclic trace property, and considering \eqref{laxeq}, one can show that,
\begin{equation}\label{Lax-conservation}
   \frac{d}{dt}\text{Tr}(L^{n})= \frac{d}{dt}H_{n}=0.
\end{equation}
 Employing the Cayley-Hamilton theorem to compute the characteristic polynomial of $L$, the coefficients in the formula are given by the elementary symmetric polynomials of the eigenvalues $\lambda_{i}$ of $L$, such that
 $$Tr(L^{n})=H_{n}=\sum_{i}\lambda^{n}_{i}.$$ 
 Using \eqref{Lax-conservation}, it is direct to show that the eigenvalues $\lambda_{i}$ are also conserved in time. When this happens, the operator $L$ is said to be \emph{isospectral} in time.

Let us assume $L$ and $M$ are defined over a Lie group $\mathcal{G}$, being $g$ an invertible element of $\mathcal{G}$, with functions on the phase space. Then \eqref{laxeq} is invariant under the simultaneous action of the similarity transformation
\begin{equation}
L=g\bar{L}g^{-1}
\end{equation}
and the gauge transformation
\begin{equation}
M=g\bar{M}g^{-1}+\frac{\partial g}{\partial t}g^{-1}.
\end{equation}

Hamiltonian system with a Lax pair structure has the benefit of ensuring the existence of infinite conserved charges. It can be proved that they are also in involution, as shown in \cite{Babelon:2003}. In this sense, and recalling the Liouville definition of integrability, their relation with Lie algebras serves as a useful manner to identify integrability in classical field theories. Indeed, recasting field equations in a Lax form will be of great usefulness in the next sections.

\section{Chern-Simons theory} \label{Section 2}
We study the canonical formulation of Chern-Simons theory in three dimensions. 
The goal of this section is to show how the dynamical content of the theory is displayed by the topological properties of the manifold and the boundary degrees of freedom captured by surface terms that make the generators of the symmetries finite. 
 We classify the gauge symmetries of the theory under the viewpoint of  ``proper",  ``improper" or  ``trivial" ones. Non-trivial boundary conditions on the fields will endow the asymptotic dynamics with the properties of integrable systems in 2D.

\subsection{The Chern-Simons action}
The Chern-Simons action for a 1-form field $A$ is given by 
\begin{equation}
I_{CS}=\frac{k}{4\pi}\intop_{M}\left\langle AdA+\frac{2}{3}A^{3}\right\rangle \,,\label{eq:Chern-Simons-form}
\end{equation}
where $M$ is some smooth 3-dimensional manifold. Here $k$ is a coupling constant that through a quantum treatment of the action, it is found to be quantized on compact manifolds\footnote{The Chern-Simons level $k$ turns out to be quantized after demanding the path integral to be gauge invariant. This has been observed for gauge transformations associated to compact groups. The Abelian Chern-Simons case serves as a model to reproduce the discrete nature of the conductivity in integer Hall effects  (see e.g. in \cite{Tong:2016kpv}).} and $A$ is a gauge field spanned in a Lie algebra $\mathfrak{g}$, $A=A_{\mu}^{I}T_{I}dx^{\mu}$ where $T_{I}$ stand for the generators $\mathfrak{g}$. The index $\mu$ labels the components associated to a certain coordinate basis. This algebra is assumed to admit an invariant non-degenerate bilinear form
$g_{IJ}=\left\langle T_{I},T_{J}\right\rangle $.
The variation of the Chern-Simons action renders the Euler-Lagrange equations
\begin{equation} \label{curvatura cero}
\mathcal{F}=dA+A^{2}=0\,,
\end{equation}
meaning the vanishing of the field strength $\mathcal{F}$. This implies that the
connection is locally flat on-shell and that there is a pure gauge solution $A=-dgg^{-1}$ for any arbitrary element $g=e^{\Lambda^{I}T_{I}}$ of the Lie algebra $\mathfrak{g}$.  

We consider the gauge transformation
\begin{equation}\label{transformacion finita}
    A=b^{-1}db+b^{-1}\bar{A}b.
\end{equation}
Infinitesimally, the group element can be expanded at linear order as $b=\mathbb{I}+\Lambda$, where $\mathbb{I}$ is the identity element, such that the gauge transformation of $A$ becomes at first order in $\Lambda$, 
\begin{equation} \label{infinitesimal transformation}
  \delta A= A-\bar{A}=d\Lambda +[A,\Lambda]. 
\end{equation}
The Chern-Simons action transforms under finite transformations \eqref{transformacion finita},
\begin{equation}
    I_{CS}[A]-I_{CS}[\bar{A}]=-\frac{k}{4\pi}\int_{\partial M}\left\langle (db\,b^{-1})\bar{A}\right\rangle-I_{\text{wn}},
\end{equation}
where 
\begin{equation}
I_{\text{wn}}=\frac{k}{12\pi}\int_{M}\left\langle (b^{-1}db)(b^{-1}db)(b^{-1}db)\right\rangle,
\end{equation}
is the winding number associated with the gauge transformation, which is a topological invariant that, for compact groups, amounts to a constant value. 
The invariance of the action principle is not ensured if the manifold is open. This is the case in the study of many field theories, where the variation of the action not only leads to terms proportional to the equations of motion but also additional surface terms that need to be handled carefully with suitable boundary conditions.
The usual choices of Dirichlet or Neumann boundary conditions can make the action principle well-defined. i.e., $\delta I=0$; nonetheless, such conditions may be too strong for many physical processes. The notion of ``edge states", or physical phenomena observed at the boundary have a crucial role on quantum gravity in the  ``holographic conjecture" \cite{Horowitz:2006ct}.

\subsection{Canonical formulation, surface terms and symmetry generators} \label{Canonical-charges-def}
The Chern-Simons action \eqref{eq:Chern-Simons-form} written
in a Hamiltonian form is
\begin{equation}
I_{H}=-\frac{k}{4\pi}\intop_{\Sigma\times R}dtd^{2}x\varepsilon^{ij}\left\langle A_{i}\dot{A}_{j}-A_{t}F_{ij}\right\rangle\,+B_{H}.\label{action Hamiltonian}
\end{equation}
This action is integrated over $M=\Sigma\times \mathbb{R}$, where $\Sigma$
is a space-like surface endowed with the coordinates $x^{i}$ where $i=1,2$ and $\mathbb{R}$ is a real time-like line parametrized by $t$.  $B_{H}$ is a boundary term included in the Chern-Simons action for having a well-defined action principle, when non-trivial boundary conditions on the gauge fields are considered, in the sense of Regge-Teitelboim \cite{Regge:1974zd} \footnote{In this notes we will restrict only to the study of surfaces terms when they are related with the generators of the symmetries.}. The 1-form connection has been split as $A=A_{i}dx^{i}+A_{t}dt$. Additionally, $\varepsilon^{12}=1$.
The Poisson bracket of the algebra-valued canonical pair $(A^{I}_{i},\pi^{Jj})$ is defined as,
\begin{equation} \label{Poisson c-s}
    \left\{A^{I}_{i}(x),\pi^{J\,j}(x') \right\}=g^{IJ}\delta^{i}_{j}\delta^{2}(x-x'),
\end{equation}
where the first term on \eqref{action Hamiltonian} is the kinetic term that determines the momenta,
\begin{equation} \label{C-S momenta}
    \pi^{i}=\frac{\partial \mathcal{L} }{\partial \dot{A}_{i}}=\frac{k}{4\pi}\epsilon^{ij}A_{j}.
\end{equation}
Here,
\begin{equation}
   \int \delta^{2}(x-x')F(x')=F(x)
\end{equation}
is the Delta-function, and $\delta^{i}_{j}$ is the Kronecker delta.
The existence of a Poisson structure  \eqref{Poisson c-s} is closely related to the existence of a symplectic structure, determined by a closed two-form field in the phase space
\begin{equation}
   \omega=\int dx^2 \langle d \pi^{i}\wedge dA_{i}\rangle,
\end{equation}
where $d$ acts similarly to an exterior derivative, but on the phase space variables.
The Dirac Poisson bracket \footnote{In here, we have used the modified Dirac Poisson bracket among the gauge fields $A$, which is non-vanishing as it has contribution of a the second class constraint \cite{Dirac,Henneaux:1992ig}.   } of any functions $F_{1}(A)$ and $F_{2}(A)$, considering \eqref{C-S momenta} and \eqref{Poisson c-s}, is defined as
\begin{equation}\label{C-S Poisson bracket}
\left\{F_{1}(A),F_{2}(A)\right\}=\frac{4\pi}{k}\int d^{2}x \frac{\delta F_{1}(A) }{\delta A^{I}_{i}}\epsilon_{ij}g^{IJ}\frac{\delta F_{2}(A) }{\delta A^{J}_{j}}.
\end{equation}

A straightforward analysis of \eqref{action Hamiltonian} shows that
$A_{t}$ is devoid of temporal derivatives, therefore, it is a Lagrange multiplier 
whereas $A_{j}$ are the dynamical fields. 
The first class constraint (in the sense of Dirac, see e.g. \cite{Dirac}), is 
\begin{equation}
G=\frac{k}{4\pi}\varepsilon^{ij}F_{ij}\,.
\end{equation}

The Hamiltonian of the theory is 
\begin{equation} \label{C-S hamiltononan}
H=-\int d^{2}x\left\langle A_{t}G\right\rangle  
\end{equation}
which vanishes for all solutions of the field equations \eqref{curvatura cero}.

The generator of the infinitesimal transformations is \eqref{infinitesimal transformation} (see e. g. \cite{PTT-Balachandran,PTT-Banados-Q,PTT-Carlip-Q}), 
\begin{equation}\label{Generator}
G\left(\Lambda\right)=\intop_{\Sigma}d^{2}x\left\langle \Lambda G\right\rangle \,.
\end{equation}
Comparing \eqref{C-S hamiltononan} and \eqref{Generator}, one can observe that the Lagrange multiplier $A_{t}$ is a subgroup of all gauge transformations generated by $\Lambda$.

To compute the infinitesimal gauge transformation on the dynamical fields one should consider \eqref{C-S Poisson bracket}, so that 
\begin{equation}\label{transformation-delta}
\delta A_{i}=\left\{ A_{i},G\left(\Lambda\right)\right\} =\frac{k}{4\pi}\left\{ A_{i},\intop_{\Sigma}d^{2}x\varepsilon^{jl}\left\langle \Lambda F_{jl}\right\rangle \right\} = \partial_{i}\Lambda+[A_{i},\Lambda].
\end{equation}
Here, the parameter $\Lambda$ is Lie algebra valued and $\Sigma$ is a closed hypersurface.

When we consider the geometry of the hypersurface to be open, e.g. the circle of infinite radius parametrized by polar coordinates $(\rho,\varphi)$, then $G\left(\Lambda\right)$ should be supplemented by a boundary term $Q\left(\Lambda\right)$ according to the Regge-Teitelboim approach \cite{Regge:1974zd}. We define the enhanced gauge generator 
\begin{equation}
\bar{G}\left(\Lambda\right)=G\left(\Lambda\right)+Q\left(\Lambda\right)\,,
\end{equation}
which improves the Hamiltonian such that its functional variation
is well-defined everywhere. Using the definition of the Poisson brackets \eqref{C-S Poisson bracket}, it is found that
\begin{equation}
\delta Q\left(\Lambda\right)=-\frac{k}{2\pi}\intop_{\rho\rightarrow\infty}\left\langle \Lambda\delta A_{\varphi}\right\rangle d\varphi .\label{eq:CanonicalGenerators}
\end{equation}
This is the functional variation of the conserved charge associated to the asymptotic gauge symmetry. It is determined by the variation of the dynamical field $A_{\varphi}$ and the gauge parameter $\Lambda$, evaluated at the boundary of $\Sigma$ on a fixed time slice.

This is a key point in the definition of the generator, as ``the problem of integrability" appears. In this context the word ``integrability" has to do with the functional integration of \eqref{eq:CanonicalGenerators}. A precise set of asymptotic conditions should be given specifying how the gauge field $A_{\varphi}$ and $\Lambda$ behave at the boundary and also their functional relation. In the next section, we will explicitly provide different boundary conditions that realize the infinite symmetries of integrable models generated by a particular kind of gauge transformation.

\subsubsection{Gauge symmetry classification}
We differentiate among different types of gauge symmetries according to the values taken by the symmetry generator $Q(\Lambda)$.

\textbf{Definition 1}:  A symmetry generated by a gauge parameter $\Lambda$ is called \textit{proper} if $Q(\Lambda)=0$.

An alternative name for them in the literature is ``small" gauge transformations, which refers to all gauge transformations that are connected to the identity element by smooth transformations. 

\textbf{Definition 2}:  A symmetry generated by a gauge parameter $\Lambda$ is called \textit{improper} if $Q(\Lambda)\neq0$.

They are also called ``large" gauge transformations, which cannot be smoothly connected with the identity.

\textbf{Definition 3}:  A symmetry generated by a gauge parameter $\Lambda$ is called \textit{trivial} if $\delta A_{\mu}=\frac{\partial I}{\partial A_{\nu}}\chi_{\mu\nu}$, where $\chi_{\mu\nu}=-\chi_{\nu\mu}$. This type of symmetry transformation trivially makes the action invariant, 
\begin{equation} \label{trivial}
  \frac{\delta I}{\delta A_{\mu}}\delta A_{\mu}=  \frac{\delta I}{\delta A_{\mu}}\frac{\partial I}{\partial A_{\nu}}\chi_{\mu\nu}=0
\end{equation}
These are not on-shell symmetries, but symmetries that vanish provided \eqref{trivial}.

\textbf{Definition 4}: In the presence of an improper symmetry, the algebra
\begin{equation}\label{symmetry-algebra}
    \left\{Q(\Lambda),Q(\bar{\Lambda})\right\}=\delta_{\bar{\Lambda}}Q(\Lambda)+ K[\Lambda,\tilde{\Lambda}]
\end{equation}
is called the \textit{asymptotic symmetry algebra}. Here, $K[\Lambda,\tilde{\Lambda}]$ is a cocyle of the algebra. The asymptotic symmetry algebra of a theory is the quotient of the algebra of allowed gauge transformations by its ideal, consisting of trivial and proper transformations.

The above definitions will be applied to Chern-Simons theories, although they are generic for any gauge theory \cite{Benguria:1976in}.

We will show how \textit{improper or ``large" gauge transformations are associated to non-trivial canonical charges}.

\subsection{Asymptotic conditions}
We consider the 1-form Chern-Simons connection to be,
\be \label{radial gaugetrans}
A=b^{-1}(\rho)(d +a)b(\rho)\,
\ee
where $b=b(\rho)$ is some group element that captures the full radial dependence, as proposed in \cite{Coussaert:1995zp}. The gauge choice \eqref{radial gaugetrans} is such that 
\be \label{boundary condition}
A|_{\rho\rightarrow\infty}\longrightarrow a,
\ee
i.e., $A$ approaches $a$ at the boundary of the manifold in a constant time slice. Then, \textit{\eqref{radial gaugetrans} represents a type of a boundary condition}, that we define as \textit{asymptotic} since it specifies the radial behavior of the field as $\rho$ tends to infinity. 

Under the same conditions \eqref{radial gaugetrans}, the 3D zero curvature equation \eqref{curvatura cero} reduces to
\begin{equation}\label{dinamica asintotica}
\partial_{t}a_{\varphi}-\partial_{\varphi}a_{t}+[a_{t},a_{\varphi}]=0,
\end{equation}
which represents the asymptotic dynamics, now given by a 2D vanishing curvature equation.

 Along this work, we consider $SL\left(2,\mathbb{R}\right)$ as the relevant symmetry group since it is crucial for $2D$ integrable systems and also for $3D$ Einstein Gravity. We define its basic elements. The corresponding Lie algebra is $\mathfrak{sl}\left(2,\mathbb{R}\right)$ that is spanned by $L_{n}$ generators, where $n\in\{-1,0,1\}$. They satisfy the commutation relation 
\begin{equation}\label{Sl(2,r)-algebra}
  \left[L_{n},L_{m}\right]=\left(n-m\right)L_{n+m}.  
\end{equation}
The Chern-Simons gauge field is spanned in the algebra $A=A^{n}L_{n}$.
We choose a matrix representation for the generators,
\begin{equation}
L_{-1}=\begin{pmatrix}0 & 0\\
1 & 0
\end{pmatrix}\quad,\quad L_{0}=\begin{pmatrix}-\frac{1}{2} & 0\\
0 & \frac{1}{2}
\end{pmatrix}\quad,\quad L_{1}=\begin{pmatrix}0 & -1\\
0 & 0
\end{pmatrix}\;.\label{PTT-sl(2,R)-MR}
\end{equation}
In this basis, the nonvanishing components of the invariant bilinear form, corresponding to the trace of the product of generators, are $\langle L_{1},L_{-1}\rangle=-1$ and $\langle L_{0}, L_{0}\rangle=1/2$.

\subsubsection{The Virasoro group and the Korteweg-de Vries   hierarchy}
We discuss some properties of the Virasoro group, which is the central extension of the diffeomorphisms in the circle $\text{Diff}(S^{1})$, commonly denoted by $\hat{\text{Diff}}(S^{1})$. The Virasoro algebra is the Lie algebra $\hat{\text{Vect}}(S^{1}) = \text{Vect}(S^{1})\oplus \mathbb{R}$, where $\text{Vect}(S^{1})$ is generated by the vector $X=X(\varphi)\partial_{\varphi}$.

The brackets of $\text{Vect}(S^{1})$ are defined by
\begin{equation}\label{DeWitt-finitealgebra}
\{\mathcal{X}(\varphi),\mathcal{Y}(\varphi)\}=\left(\mathcal{X}(\varphi)\partial_{\varphi}\mathcal{Y}(\varphi)-\mathcal{Y}(\varphi)\partial_{\varphi}\mathcal{X}(\varphi)\right)  \partial_{\varphi}
\end{equation}
Since $\mathcal{X},\mathcal{Y}$ are functions on the circle, they can be expanded in Fourier modes as,
\begin{equation}
    \mathcal{X}=\sum_{n\in\mathbb{Z}} \mathcal{X}_{n}l_{n}\quad\text{where}\quad l_{n}=e^{in\varphi} \partial_{\varphi}.
\end{equation}
The brackets \eqref{DeWitt-finitealgebra} leads to the Witt algebra, 
\begin{equation}
   i\{l_{n},l_{m}\}=(n-m)l_{n+m}
\end{equation}
which is infinitely generated, as $n,m\in\mathbb{Z}$. This is the algebra of infinitesimal conformal transformations in an Euclidean two-dimensional flat space. The modes $n\in\{-1,0,1\}$ generate
$\mathfrak{sl}\left(2,\mathbb{R}\right)$ as a subalgebra. 

We define now the generators $\mathcal{L}(t,\varphi)$ as the elements of the Virasoro algebra, whose finite commutation relation at equal time is 
\begin{eqnarray}\label{Virasoro-algebra}
\{\mathcal{L}(\varphi),\mathcal{L}(\bar{\varphi})\}=(2\mathcal{L}(\varphi)\partial_{\varphi}+\partial_{\varphi}\mathcal{L}(\varphi)-c\partial^{3}_{\varphi})\delta(\varphi-\bar{\varphi})    
\end{eqnarray}
Here $c$ is the central charge \footnote{Central charges are non-trivial two-cocycles in the canonical realization of the asymptotic symmetry algebra \eqref{symmetry-algebra}.}. 
There is an infinite non-linear arrangement,
\begin{eqnarray}\label{cargas kdv}
 H_0&=&\int d\varphi\,\cL,\qquad
 H_1= \int d\varphi\, \cL^2\\
 H_2&=& \int d\varphi\, \left(\cL^3+c  \cL'^2\right),\\
 H_3&=& \int d\varphi \, \left(\cL^4+4c\cL\cL'^2+\frac{4}{5}c^{2}\cL''^{2}\right),\quad\cdots.
\end{eqnarray}
that leads to an infinite set of ``polynomial" functions, namely, the Gelfand-Dikkii polynomials (for a review on KdV see \cite{Miura}). From now one, we use interchangeably the notation $\partial_{\varphi}(\,)=(\,)'$ and $\partial_{t}(\,)=\dot{(\,)}$. Provided \eqref{Virasoro-algebra}, the integrals \eqref{cargas kdv} satisfy a commuting algebra,
\begin{equation}
 \{H_{i},H_{j}\}=0.   
\end{equation}
They are the infinite conserved currents of the chiral equation $\cL'=\dot{\cL}$ and also, of the Korteweg de Vries (KdV) equation, 
    \begin{equation} \label{KdV equation}
\dot{\cL}= 3\cL'\cL - c \cL '''.
\end{equation}
In fact, the chiral equation and the KdV equation belong to an infinite set of non-linear differential equations that share the same tower of conserved charges. Together, they are named as the KdV hierarchy. 
We will construct the infinite tower of charges \eqref{cargas kdv}, from improper gauge symmetries of the Chern-Simons generators. We will also construct this infinite tower of non-linear KdV hierarchy equations from its asymptotic dynamics.

\subsubsection*{KdV asymptotic conditions}
As motivated in \eqref{radial gaugetrans} and considering the choice of Coussaert, Henneaux and Van Driel \cite{Coussaert:1995zp}, we propose $b=\exp[\rho L_0]$ as a suitable group element $b$ that ensures \eqref{boundary condition} as $\rho\rightarrow\infty$. The connection components take the form

\begin{eqnarray} \label{eq:kdvbc}
a_{\varphi}& =&- \frac{2\pi}{k}\cL(t,\varphi)L_{-1}+L_{1}-2\lambda L_{0},\\ \label{bc_aphi}
a_{t}& =& \epsilon  L_{1}- \left(\frac{\epsilon '}{2}+ \lambda  \epsilon \right) L_{0}+\left(-\frac{\epsilon ''}{2}-\lambda  \epsilon '+\frac{2 \pi  \epsilon \mathcal{L}}{k} \right)L_{-1}\,,\label{bc_at}
\end{eqnarray}
where $\lambda$ is a constant without variation and $\{L_{-1},L_{0},L_{1}\}$ are the generators of $\mathfrak{sl}(2,\mathbb{R})$ defined in \eqref{PTT-sl(2,R)-MR}.  The asymptotic zero curvature equation \eqref{dinamica asintotica} reduces to,
 \begin{equation} \label{KdV hierarchy}
\dot{\cL}=\epsilon  \cL' +2\cL \epsilon ' - \frac{k}{4\pi} \epsilon '''+\frac{k}{\pi}\lambda^2 \epsilon ' .    
\end{equation}
A similar setup to study KdV can be found in, e.g., \cite{Babelon:2003}; nonetheless, in this manuscript, the spirit is different because there is a gauge choice that permits the reduction of the dynamics of a $2+1$ system to one dimension less.
It has been considered that extending the gauge group to complex variables, i.e. to $SL(2,\mathbb{C})$, one can take the function $\epsilon$ to be expanded in Laurent series of $\lambda$, that introduces pole singularities in the flat gauge connections. This is the Zakharov–Shabat construction \cite{Zakharov:1979zz}, where matrices depend on some spectral parameter as rational functions \footnote{The reason to call $\lambda$ \textit{spectral parameter} has to do with its role in the auxiliary linear problem because they are eigenvalues of the associated Schr\"odinger equation. This will be shown in Section \ref{Section Flat}.}. In the context of the Inverse Scattering problem \cite{book}, a similar expansion was considered by Ablowitz. \textit{et. al.} \cite{Ablowitz:1974ry}. 
As we will see here, it permits the construction of a collection of integrable systems. 

We present the above claim explicitly for the particular example of KdV, where we restrict to real components of the gauge fields, as it is sufficient for this case. The function $\epsilon$ in the the Lagrange multiplier $a_{t}$, is expanded in even powers of $\lambda$, 
\begin{equation}\label{expansion}
\epsilon=\sum_{n=0}^{N}\epsilon_{n}\lambda^{2(N-n)},
\end{equation}
where $N$ is an arbitrary positive integer. 
The field equation for $\cL$ is obtained considering the $\lambda^0$ term
\begin{equation}
\dot{\cL}=\epsilon_{N} \cL' +2\cL \epsilon_{N}' - \frac{k}{4\pi} \epsilon_{N}'''.
\end{equation}\label{eomKDV}
We introduce the  operator
$$\mathcal{D}= \cL' +2\cL \partial_{\varphi} - \frac{k}{4\pi} \partial^{3}_{\varphi}$$
where one can recognize the differential symplectic operator of the Virasoro algebra \eqref{Virasoro-algebra}, for $c=\frac{k}{4\pi}$. The remaining equations imply a recursion relation for the coefficients in the expansion
\begin{equation}\label{recurrence kdv}
 \epsilon'_{n+1}=-\frac{4\pi}{k}\mathcal{D}\epsilon_{n}\quad\text{for}\quad 0\leq n \leq N-1
\end{equation}
with the additional condition $\epsilon'_{0}=0$. This recursion relation permits to construct an infinite set of differential equations, the KdV hierarchy, all sharing the same infinite tower of conserved currents.

In the words of Dickey in \cite{Dickey:1993ih}, we provide a definition for a hierarchy: 

``Which property permits to consider a hierarchy as a single whole, as an entity? Geometrically speaking, a differential equation is a flow in a phase space. We call a family of differential equations a hierarchy if they act in the same phase space and commute. Then each of equations determines a symmetry for each other. Let us consider this as a characteristic property of a hierarchy."

The dynamics of the hierarchy can be written as Dickey states, meaning a differential flow in phase space
\begin{equation}
   \dot{\cL}=\mathcal{D}\epsilon_{N}=\mathcal{D}\left(\frac{\delta H_{N} }{\delta \cL}\right).
\end{equation}

The choice of $N$ solves a functional form of $\epsilon_{N}$ in \eqref{recurrence}, that renders the corresponding differential equation of the hierarchy. For example, taking $N=0$, one obtains the condition $\epsilon_{0}=cte$. Applying this value to the equation for the asymptotic dynamics \eqref{eomKDV}, one finds $\cL'=\dot{\cL}$. Taking now $N=1$, the recurrence \eqref{recurrence} leads to,
\begin{equation}
\epsilon_1=-\frac{4\pi}{k}\epsilon_{0} \cL + \bar{\epsilon}_1,
\end{equation}
where $\epsilon_{0},\bar{\epsilon}_1$ are constants associated to integrating the recurrence relation. 
The asymptotic dynamics reduce
\begin{equation} \label{kdveq}
    \dot{\cL}=3\cL\cL'-\frac{k}{4\pi}\cL'''-\frac{k\, \bar{\epsilon}_1}{4\pi\,\epsilon_{0}}\cL'.
\end{equation}\
We retrieve the KdV equation choosing $\bar{\epsilon}_1=0$.
Higher values of $N$ lead to other higher-order non-linear differential equations of the hierarchy.

\subsubsection*{Canonical charges}
Coming back to subsection \ref{Canonical-charges-def}, we observe that from the comparison of equations \eqref{C-S hamiltononan} and \eqref{Generator}, the Lagrange multiplier $a_{t}$ is a subset of all gauge transformations generated by $\Lambda$. To recover the infinite tower of conserved charges as canonical generators \eqref{eq:CanonicalGenerators}, one chooses this relation to propose a gauge generator $\Lambda$ with the same functional form as $a_{t}$ in \eqref{bc_at}. Indeed, we set
\begin{equation} \label{Lambda KdV}
    \Lambda=\mu  L_{1}- \left(\frac{\mu '}{2}+ \lambda  \mu \right) L_{0}+\left(-\frac{\mu ''}{2}-\lambda  \mu '+\frac{2 \pi \mu\mathcal{L}}{k} \right)L_{-1}\,,
\end{equation}
with $\mu=\mu(t,\varphi)$. We now study the infinitesimal gauge transformation
\begin{equation}\label{angular-gauge-transformation}
\delta a_{\varphi}=\partial_{\varphi}\Lambda+[a_{\varphi},\Lambda],
\end{equation}
demanding this equation to preserve the boundary conditions \eqref{eq:kdvbc}, i.e.  we look for symmetry transformations mapping the fields into themselves,
\be \label{symmetry-transformation}
\delta a=\cO(a)\,.
\ee
The notation $\mathcal{O}(\text{field})$ on the right hand side of the equation,  implies that the gauge transformation acts on the field,  in this case $a$,  such that its form is preserved up to redefinitions $a \rightarrow a+\delta a $.  In the particular case given in \eqref{eq:kdvbc},  this implies a transformation of the function $\mathcal{L} \rightarrow \mathcal{L}+\delta \mathcal{L}$.  Demanding \eqref{angular-gauge-transformation} and \eqref{symmetry-transformation},  one finds what is called the \textit{symmetry transformation}, 
 \begin{equation} \label{KdV transformation}
\delta\cL=\mu \cL' +2\cL \mu ' - \frac{k}{4\pi} \mu '''+\frac{k}{\pi}\lambda^2 \mu ' .    
\end{equation}
Assuming that $\mu$ is chosen according to the same expansion \eqref{expansion},
\begin{equation} \label{recurrence gauge}
    \mu=\sum_{n=0}^{N}\mu_{n}\lambda^{2(N-n)},
\end{equation}
then the two conditions \eqref{KdV transformation} and \eqref{recurrence gauge} recover the same previous recurrence \eqref{recurrence kdv}, now for the gauge parameter $\mu_{n}$,
\begin{equation}\label{recurrence kdv-2}
 \mu'_{n+1}=-\frac{4\pi}{k}\mathcal{D}\mu_{n}\quad\text{for}\quad 0\leq n \leq N-1
\end{equation}
We evaluate this result on the canonical generator \eqref{eq:CanonicalGenerators}. Considering \eqref{Lambda KdV} and \eqref{bc_aphi}, it leads to
\begin{equation} \label{kdv carga}
\delta Q\left(\Lambda\right)=-\int \mu\delta \cL  d\varphi. 
\end{equation}
The remaining task is ``to take the $\delta$ away". By solving the recurrence \eqref{recurrence kdv-2} iteratively, one finds the explicit value for each element of the expansion, which permits the evaluation and integration of the canonical generator \eqref{eq:CanonicalGenerators}.
We give two examples to show how this happens using \eqref{recurrence kdv-2}. For $N=0$, $\mu_{0}=cte$ that we chose equal to $-1$. For $N=1$, there are two integration constants, and we fix them so that $\mu_{1}=-\frac{\cL}{2}$. We use them individually to integrate in phase space \eqref{eq:CanonicalGenerators}, that gives the first two integrals in \eqref{cargas kdv} for particular values of its functions.

In general, \eqref{kdv carga} reduces to 
\begin{equation} \label{canonical-kdv}
    Q(\Lambda)=\sum_{n=0}^{N}\zeta_{n}H_{n}.
\end{equation}
Here $H_{n}$ is the $n-$th element of the KdV tower of conserved charges \eqref{cargas kdv}, associated to $\mu_{n}$, with $c=\frac{k}{4\pi}$ and $\zeta_{n}$ arbitrary constants. 
The involution of these charges will be shown in the next subsection, which will devoted to the AKNS integrable system. AKNS integrable system contains the KdV hierarchy. 

\subsubsection{AKNS boundary conditions}
 The Ablowitz-Kaup-Newell-Segur (AKNS)  integrable hierarchy \cite{Ablowitz:1973zz} can also be obtained as the asymptotic dynamics of Chern-Simons field equations.
 Following Ref. \cite{Ablowitz:1974ry, Cardenas:2021vwo, Cardenas:2025qqi}, we propose that the angular component in \eqref{radial gaugetrans} reads as follows,
\begin{equation}\label{aphi}
a_{\varphi}=- 2\xi L_{0}-p L_{1}+r L_{-1}.
\end{equation}
The component along $L_0$ is chosen to be a constant without variation at the boundary. The temporal component of the gauge connection is given by
\begin{equation}\label{at}
a_{t}=-2A L_{0}+ B L_{1} -C L_{1}.
\end{equation}
We find that the vanishing of the curvature two-form \eqref{curvatura cero} coincides with AKNS integrable equations, 
\begin{subequations}\label{aknss}
  \begin{align}\label{akns1}
    \dot{r}+\left(C'-2r A-2\xi C\right)&=0 \, , \\ \label{aknss2}
    \dot{p}+ \left( B'+2p A+2\xi B\right)&=0\, , \\ \label{aknss3}
    A'-p C+r B&=0 \, .
  \end{align}
\end{subequations}

To solve them, we repeat the same strategy followed in the KdV case and propose $A$, $B$, and $C$ to be constructed recursively. They are polynomials in $\xi$ with coefficients $A_{n}$, $B_{n}$, and $C_{n}$ depending on $p$, $r$, and its derivatives, expanded as,
\begin{equation}\label{rec_sol}
    A=\sum_{n=0}^{N} A_{n}\xi^{N-n},\,\, B=\sum_{n=0}^{N} B_{n}\xi^{N-n}, \,\, C=\sum_{n=0}^{N} C_{n}\xi^{N-n},
\end{equation}
where $N$ is an arbitrary positive integer. The $\xi^0$ terms in \eqref{aknss} gives the dynamical equations
\begin{equation}\label{eomN}
 \dot r = \frac{1}{\ell}\left(-C_N'+2 r A_N\right),\quad  \dot p = \frac{1}{\ell}\left(-B_N'-2 p A_N\right).
\end{equation}
The remaining terms imply the recursion relations for the coefficients in the expansion
\begin{subequations}\label{recurrence}
  \begin{align}
    A_{n}'&=pC_{n}-rB_{n}, \label{recurrenceA}\\
    B_{n+1}&=-\frac{1}{2}B_{n}'-p A_{n},  \\
    C_{n+1}&=\frac{1}{2}C_{n}'-r A_{n},
  \end{align}
\end{subequations}
along with $B_0=C_0=0$, which can be solved for each coefficient \cite{trace}. 
As in the example with KdV boundary conditions, different values of $N$ lead to distinct asymptotic dynamics. Nonetheless, in this case \emph{the AKNS hierarchy encompasses a richer structure}, as it contains several well-known integrable equations (and their hierarchies), that arise as particular cases of the above construction: Korteweg-de Vries ($N=3$, $r=1$), modified Korteweg-de Vries ($N=3$, $r=p$), (Wick rotated) nonlinear Schr\"{o}dinger ($N=2$), chiral boson ($N=1$), among others. The Sine-Gordon equation is also included taking negative powers of $\xi$ and considering $r=-p$. 

\subsubsection*{Conserved charges}

The asymptotic symmetries are constructed from improper transformations associated to 
\begin{equation} \label{gauge-trans}
\delta a = d \Lambda + \left[ a,\Lambda \right],
\end{equation}
where we respect the form of the boundary conditions along the temporal component \eqref{at} to construct the generator $\Lambda$. 
We consider the general gauge parameter 
\begin{equation} \label{gauge-param-akns}
     \Lambda=-2\alpha L_{0}+\beta L_{1}-\gamma L_{- 1},
\end{equation}
and assume an expansion in $\xi$ for $\alpha$, $\beta$, and $\gamma$ given by,
\begin{subequations} \label{gauge-param}
\begin{align}
    &\alpha=\sum_{n=0}^{M} \frac{(n-1)}{2}\mathcal{H}_n\xi^{M-n},\,\,  \\
    &\beta=\sum_{n=0}^{M}\mathcal{R}_{n+1}\xi^{M-n}, \,\,\\
    & \gamma=\sum_{n=0}^{M} \mathcal{P}_{n+1}\xi^{M-n},
\end{align}
\end{subequations}
where 
\begin{equation} \label{rec_solution2}
    \alpha_n=\frac{n-1}{2}\mathcal{H}_n,\quad \beta_n=\mathcal{R}_{n+1},\quad \gamma_n=\mathcal{P}_{n+1},
  \end{equation}
for $n \geq 1$. Here $M$ is a positive integer that labels the infinite family of permissible gauge transformations. The quantities $\mathcal{R}_{n}$ and $\mathcal{P}_{n}$ correspond to variational derivatives of a function $H_n[r,p]$ in the following way,   
\begin{equation} \label{functiontal-rel}
\mathcal{R}_n = \frac{\delta H_n}{\delta r} , \quad \mathcal{P}_n =\frac{\delta H_n}{\delta p},\quad\text{where} \quad H_n=\int \mathcal{H}_n \,d\varphi.
\end{equation}  
The asymptotic symmetry transformations of the fields $r$ and $p$ are then given by 
\begin{equation}\label{eomM}
  \delta r = -\gamma_M'+2 r \alpha_M,\quad \delta p = -\beta_M'-2 p \alpha_M.
\end{equation}

In the canonical formulation, we construct conserved charges through the boundary term $Q[\Lambda]$, of which we only have its functional variation \eqref{eq:CanonicalGenerators}. For the proposed boundary conditions \eqref{aphi} and \eqref{gauge-param-akns}, it is given by 
\begin{equation}\label{deltaQ}
\delta Q[\Lambda]=\frac{k}{2\pi} \int d\varphi \left( \beta \delta r+\gamma \delta p\right),
\end{equation}
By virtue of equations \eqref{gauge-param} and \eqref{functiontal-rel}, it can be integrated to
\begin{equation}\label{Q}
 Q[\Lambda] =\frac{k}{2 \pi} \sum_{n=0}^M \zeta_{M-n}H_{n+1}.
\end{equation}
where $\zeta_{M-n}$ are arbitrary constants and
\begin{equation}\label{AKNS-Cargas}
 H_{2}=-\int pr \,d\varphi,\quad H_{3}=\frac{1}{4}\int (p'r-pr')d\varphi,\quad H_{4}=\frac{1}{4}\int (p^{2}r^{2}+p'r')d\varphi,\quad...
\end{equation}
 while $H_{1}=0$. 
\subsubsection*{Bi-Hamiltonian structure and charges in involution}

Another important feature of integrable systems relies on its bi-Hamiltonian structure. A bi-Hamiltonian system is characterized by two symplectic operators, that imply two different Poisson brackets generating the same dynamics for a given integrable system. Let us consider the Hamiltonian operators\footnote{We define the antiderivative $\partial_{\phi}^{-1}f(\phi)=\int_{\infty}^{\phi}\left(f(y)\right)dy$, where we set the value of the integrand to be zero at infinity.}
\begin{equation}\label{simplectic1}
\mathcal{D}_{1}=\left(\begin{array}{cc}
-2r\partial_{\varphi}^{-1}\left(r\cdot\right) & \quad-\partial_{\varphi}+2r\partial_{\varphi}^{-1}\left(p\cdot\right)\\
-\partial_{\varphi}+2r\partial_{\varphi}^{-1}\left(p\cdot\right) & 2p\partial_{\varphi}^{-1}\left(p\cdot\right)
\end{array}\right)
\end{equation}
and 
\begin{equation}\label{simplectic2}
\mathcal{D}_{2}=\left(\begin{array}{cc}
0 & \quad-2\\
2 & 0
\end{array}\right).
\end{equation}

In the AKNS case, we have that their equations of motion \eqref{aknss} can be written in the following manner, 
\begin{equation}
\left(\begin{array}{c}
\dot{r}\\
\dot{p}
\end{array}\right)=\mathcal{D}_{1}\left(\begin{array}{c}
\mathcal{R}_{N+1}\\
\mathcal{P}_{N+1}
\end{array}\right)=\mathcal{D}_{2}\left(\begin{array}{c}
\mathcal{R}_{N+2}\\
\mathcal{P}_{N+2}
\end{array}\right).
\end{equation}

Moreover, such operators $\mathcal{D}$ permits to write Poisson brackets of two arbitrary functionals $F=F\left[r,p\right]$ and $G=G[r,p]$,
\begin{equation}\label{AKNS-Poisson}
\left\{ F\left(\phi\right),G\left(\phi'\right)\right\} =\int d\varphi\left(\begin{array}{cc}
\frac{\delta F\left(\phi\right)}{\delta r\left(\varphi\right)} & \frac{\delta F\left(\phi\right)}{\delta p\left(\varphi\right)}\end{array}\right)\mathcal{D}\left(\begin{array}{c}
\frac{\delta G\left(\phi'\right)}{\delta r\left(\varphi\right)}\\
\frac{\delta G\left(\phi'\right)}{\delta p\left(\varphi\right)}
\end{array}\right).
\end{equation}

The operators $\mathcal{D}_1$ and $\mathcal{D}_2$ satisfy the property of compatibility, which means that $  \mathcal{D}_1+\mathcal{D}_2$ is also a Hamiltonian operator \cite{Olver}.
On the other hand, the recurrence relation \eqref{recurrence}, can be rewritten using both operators as
\begin{equation}\label{bihamrec}
  \left(
    \begin{aligned}
      \mathcal{R}_{n+1}\\ \mathcal{P}_{n+1}
    \end{aligned}
\right) = \mathcal{D}_2^{-1}\mathcal{D}_1 \left(
    \begin{aligned}
      \mathcal{R}_{n}\\ \mathcal{P}_{n}
    \end{aligned}
\right).
\end{equation}

From the Liouville perspective of integrability,  and extending this notion to field theory,  an integrable system is endowed with a infinite set of charges \eqref{AKNS-Cargas}  that commute with the Hamiltonian and among themselves. A manner to prove the existence of an Abelian charge algebra is using the properties of its bi-Hamiltonian structure.

%From the Liouville perspective of integrability,  one can extended this notion to field theory,  such that

%Now we are in position to show the generators \eqref{Q}  are in involution, $$\{Q[\Lambda],Q[\bar \Lambda]\}=0,$$
%where $\Lambda$ and $\bar \Lambda$ stand for two different gauge parameters generating any conserved current of the tower \eqref{AKNS-Cargas}. The commutation relation can be computed using the canonical Poisson bracket \eqref{AKNS-Poisson}. 

\textit{Proof:}\, Following \cite{Brown:1986ed}, we consider the equation,
\begin{equation}\label{charges-comm}
   \{Q[\Lambda],Q[\bar \Lambda]\}=\bar \delta Q[\Lambda], 
\end{equation}
which means that charges are canonical generators of infinitesimal transformations. Here $\bar \delta$ acts as the gauge transformation \eqref{transformation-delta} generated by $\bar \Lambda$.  

Let us consider the variation of our canonical charge \eqref{deltaQ}.  Additionally, let us note that the field variations \eqref{eomM} can be written in terms of the symplectic operators \eqref{simplectic1} and  \eqref{simplectic2} as, 
\begin{equation}
\left(\begin{array}{c}
\delta{r}\\
\delta{p}
\end{array}\right)=\mathcal{D}_{1}\left(\begin{array}{c}
\mathcal{R}_{M+1}\\
\mathcal{P}_{M+1}
\end{array}\right)=\mathcal{D}_{2}\left(\begin{array}{c}
\mathcal{R}_{M+2}\\
\mathcal{P}_{M+2}
\end{array}\right) , 
\end{equation}
so that the right hand side of the charge algebra \eqref{deltaQ} is given by
\begin{equation} \label{charge proof 1}
\{Q[\Lambda],Q[\bar \Lambda]\}=\frac{k\ell}{2\pi}  \sum_{n=0}^M \lambda^{M-n}\int d\varphi
\begin{pmatrix}
  \mathcal{R}_{n+1} & \mathcal{P}_{n+1}
\end{pmatrix}
\mathcal{D}_1
\begin{pmatrix}
  \mathcal{R}_{\bar M +1} \\ \mathcal{P}_{\bar M +1}
\end{pmatrix}.
\end{equation}
We use now the property of antisymmetry of the Poisson brackets, which permits to rewrite the above as
\begin{equation}
\{Q[\Lambda],Q[\bar \Lambda]\}=-\frac{k\ell}{2\pi}  \sum_{n=0}^M \lambda^{M-n}\int d\varphi
\begin{pmatrix}
  \mathcal{R}_{\bar M+1} & \mathcal{P}_{\bar M+1}
\end{pmatrix}
\mathcal{D}_1 \begin{pmatrix}
  \mathcal{R}_{ n +1} \\ \mathcal{P}_{ n +1}
\end{pmatrix}.
\end{equation}
Here, we recall that the symplectic operator acts as follows, 
\begin{equation}
\mathcal{D}_1  \begin{pmatrix}
  \mathcal{R}_{ n +1} \\ \mathcal{P}_{ n +1}
\end{pmatrix}=\mathcal{D}_2  \begin{pmatrix}
  \mathcal{R}_{ n +2} \\ \mathcal{P}_{ n +2}
\end{pmatrix}.
\end{equation}
so that, 
\begin{align}
\{Q[\Lambda],Q[\bar \Lambda]\}=-\frac{k\ell}{2\pi}  \sum_{n=0}^M \lambda^{M-n}\int d\varphi
\begin{pmatrix}
  \mathcal{R}_{\bar M+1} & \mathcal{P}_{\bar M+1}
\end{pmatrix}
\mathcal{D}_2 \begin{pmatrix}
  \mathcal{R}_{ n +2} \\ \mathcal{P}_{ n +2}
\end{pmatrix} \\
=\frac{k\ell}{2\pi}  \sum_{n=0}^M \lambda^{M-n}\int d\varphi
\begin{pmatrix}
  \mathcal{R}_{n+2} & \mathcal{P}_{ n+2}
\end{pmatrix}
\mathcal{D}_2 \begin{pmatrix}
  \mathcal{R}_{ \bar M+1} \\ \mathcal{P}_{ \bar M+1}
\end{pmatrix}\\
=\frac{k\ell}{2\pi}  \sum_{n=0}^M \lambda^{M-n}\int d\varphi
\begin{pmatrix}
  \mathcal{R}_{n+2} & \mathcal{P}_{ n+2}
\end{pmatrix}
\mathcal{D}_1 \begin{pmatrix}
  \mathcal{R}_{ \bar M} \\ \mathcal{P}_{ \bar M}
\end{pmatrix}  \label{charge proof2}
\end{align}

Indeed, one can always apply $k$ times the above procedure, so that 
\begin{equation*}
 \sum_{n=0}^M \lambda^{M-n}\int d\varphi
\begin{pmatrix}
  \mathcal{R}_{n+1} & \mathcal{P}_{n+1}
\end{pmatrix}
\mathcal{D}_1
\begin{pmatrix}
  \mathcal{R}_{\bar M +1} \\ \mathcal{P}_{\bar M +1}
\end{pmatrix}= \sum_{n=0}^M \lambda^{M-n}\int d\varphi
\begin{pmatrix}
  \mathcal{R}_{n+1+k} & \mathcal{P}_{n+1+k}
\end{pmatrix}
\mathcal{D}_1
\begin{pmatrix}
  \mathcal{R}_{\bar M +1-k} \\ \mathcal{P}_{\bar M +1-k}
\end{pmatrix}
\end{equation*}
Since $k$ is an arbitrary number, one can choose it so that  
so that $n+k=\bar M$ , that imply $k=\bar M -n$ and 

\begin{eqnarray} 
\{Q[\Lambda],Q[\bar \Lambda]\}&=&\frac{k\ell}{2\pi}  \sum_{n=0}^M \lambda^{M-n}\int d\varphi
\begin{pmatrix}
  \mathcal{R}_{\bar M +1} & \mathcal{P}_{\bar M +1}
\end{pmatrix}
\mathcal{D}_1
\begin{pmatrix}
  \mathcal{R}_{n +1} \\ \mathcal{P}_{n +1}
\end{pmatrix}\\
&&=\{Q[\bar\Lambda],Q[ \Lambda]\}
\end{eqnarray}

Which completes the proof. The generators of this family are in involution, i.e., span an Abelian algebra 
$$\{Q[\Lambda],Q[\bar \Lambda]\}=0.$$

\section{Applications in 3D Gravity} \label{Section 3}
The richness of Einstein gravity with negative cosmological constant in three dimensions is strongly related with its asymptotic symmetry group; indeed, it is larger than the isometries of the spacetime $SO(2,2)$, and is given by the conformal group in two dimensions \cite{BrownHenneaux}, that it is infinitely generated. This means that using the AdS/CFT
correspondence \cite{Maldacena:1997re}, this conformal field theory should correspond to the quantum AdS gravity in three dimensions. Additionally, the existence of a black hole solution, the Bañados-Teitelboim-Zanelli (BTZ) black hole \cite{Banados:1992wn,Banados:1992gq}, has served as a laboratory to test quantum issues \cite{Strominger:1997eq,Carlip:2004ba}. Also, the formulation of the theory in terms of a Chern-Simons action and the further understanding on how to implement asymptotic conditions from that formalism   \cite{Coussaert:1995zp,Chemical pot,Bunster:2014mua}, has helped to promote a larger advance and find new asymptotic symmetries (see e.g.\cite{Afshar:2016kjj,Grumiller:2016pqb,Grumiller:2017sjh,Henneaux:2019sjx,Mertens:2022ujr}).

Here we focus on the latest progress that study the asymptotic symmetries of gravity in connection to integrable systems \cite{Perez:2016vqo,Grumiller:2019tyl,Fuentealba:2017omf,Gonzalez:2018jgp,Melnikov:2018fhb,Ojeda:2019xih,Erices:2019onl,Ojeda:2020bgz,Dymarsky:2020tjh,Cardenas:2021vwo,Lara:2024cie}, and present as an example, new asymptotic conditions that realize the AKNS asymptotic dynamics.

\subsection{Relationship with General Relativity in 3D dimensions}

The Einstein-Hilbert action principle for gravity with a negative cosmological constant $\Lambda=-\frac{1}{\ell^{2}}$, can be represented by the difference of two Chern-Simons
actions \footnote{Up to some boundary terms that we will not deal with along this notes.} on the $SL(2,\mathbb{R})$ group \cite{Achucarro:1987vz,Witten:1988hc}
\begin{align}
I_{E} & =\frac{k}{4\pi}\int\sqrt{g}\left(R+\frac{2}{\ell^{2}}\right)=I_{CS}\left[A^{+}\right]-I_{CS}\left[A^{+}\right]
\end{align}
taking
\begin{equation}
\mathcal{A}^{\pm}=\left(\text{\ensuremath{\omega}}^{a}\pm\frac{e^{a}}{\ell}\right).\label{eq:cs-conection}
\end{equation}
Here the field $e^{a}=e_{\mu}^{a}dx^{\mu}$ is the dreibein 1-form
and $\omega^{a}=\omega_{\mu}^{a}dx^{\mu}$ is the dualized spin connection
1-form, which define the dualized curvature 2-form $R^{a}=d\omega^{a}+\frac{1}{2}\epsilon^{abc}\omega_{b}\omega_{c}$ and the torsion $T^{a}=de^{a}+\epsilon^{abc}\omega_{b}e_{c}$ in the first order formulation of the theory. For more details, see, for example, \cite{Carlip:2004ba}.
Each ``left'' (-) and ``right'' (+) sector is spanned by the $\mathfrak{sl}(2,\mathbb{R})$ algebra. This is because the isometries of $AdS_{3}$ can be seen
as the direct sum 
\begin{equation}
SL(2,\mathbb{R})^{+}\oplus SL(2,\mathbb{R})^{-}
\end{equation}
This feature is particular to $AdS_{3}$, and we take advantage of this group structure and the formulation of the theory as a gauge theory to replicate in both sectors the integrable asymptotic dynamics previously discussed. 
 Einstein's equations, in turn, are equivalent to the zero curvature conditions $\mathcal{F}^{\pm}=0$, where $\mathcal{F}^{\pm}=d\mathcal{A}^\pm+\mathcal{A}^\pm \wedge \mathcal{A}^\pm$. The metric field can be constructed from the gauge fields as 
\begin{equation}\label{gmunu}
  g_{\mu\nu}=\frac{\ell^2}{2}\left\langle \left(\mathcal{A}_{\mu}^{+}-\mathcal{A}_{\mu}^{-}\right),\left(\mathcal{A}_{\nu}^{+}-\mathcal{A}_{\nu}^{-}\right)\right\rangle, 
\end{equation}
where $\langle\, , \, \rangle$ is the invariant bilinear form of the gauge group.

As presented in \cite{Coussaert:1995zp}, the boundary conditions  for the gravitational case comprise all the gauge fields of the form ,
\begin{equation}\label{bdab}
\mathcal{A}^\pm=b^{\mp 1}( d+ a^\pm ) b^{\pm 1},
\end{equation}
where the connections $a^\pm=a^{\pm}_{\varphi}d\varphi+a^{\pm}_{t}dt$ depend only on $t$ and $\varphi$. Hence, the gauge group element $b\left(\rho\right)$ completely captures the radial dependence of the fields. 

\subsection{Metric in Gaussian normal
coordinates}

Near horizon metrics have been recently studied to look for non-trivial symmetry groups at the black hole horizon (as constructed, e.g, in \cite{Donnay:2015abr,Grumiller:2019fmp} or in\cite{
Grumiller:2019tyl,Ojeda:2019xih,Cardenas:2021sun}, using the Chern-Simons formulation). It has been proposed that enlarged symmetries could provide an explanation to the information loss paradox \cite{Hawking,Hawking:2016msc}. Here, following \cite{Afshar:2016kjj}, we recover a geometry of this type by restoring the radial coordinate through the gauge transformation,
\begin{equation}
b_{\pm}=\exp\left(\pm\frac{\rho}{2\ell}L_{0}\right).
\end{equation}
In a geometric sense, by near horizon, it is implied that there is an expansion of the metric coefficients around $\rho=0$ that corresponds, in the stationary case, to black hole solutions that are not necessarily spherically symmetric and whose horizon is located at that value of the radial coordinate. In this case, the asymptotic conditions specify the radial behavior of the metric components as it approaches the hypersurface $\rho=0$.

The metric components are recovered considering the form of the gauge fields as in \eqref{bdab} and the formula that relates gauge fields and the spacetime metric \eqref{gmunu}. It permits the construction of the following components,
\begin{eqnarray*}
g_{tt} & =&(A^{-}-A^{+})^{2}-(B^{-}-C^{+})(B^{+}-C^{-})+(C^{+}C^{-}-B^{+}B^{-})\frac{\rho}{\ell}\\
&&-(C^{+}C^{-}+B^{+}B^{-})\frac{\rho^{2}}{\ell^{2}}+\mathcal{O}\left(\rho^{3}\right)\\
g_{t\rho} & =&\frac{1}{2}(A^{-}-A^{+})+\mathcal{O}\left(\rho\right)^{2},\\
g_{t\phi} & =&\ell(\xi^{+}+\xi^{-})(A^{+}-A^{-})+\frac{\ell}{2}(p^{-}+r^{+})\left(C^{-}-B^{+}\right)-\frac{\ell}{2}(p^{+}+r^{-})\left(C^{+}-B^{-}\right)\\
 && +\frac{\rho}{2}(B^{-}p^{+}-B^{+}p^{-}+C^{+}r^{-}-C^{-}r^{+})+\frac{\rho^{2}}{4\ell}(B^{-}p^{+}-B^{+}p^{-}-C^{+}r^{-}+C^{-}r^{+})+\mathcal{O}\left(\rho^{3}\right)\\
g_{\rho\rho} & =&\frac{1}{4}+\mathcal{O}\left(\rho\right)^{2},\\
g_{\rho\phi} & =&-\frac{\ell}{2}(\xi^{+}+\xi^{-})+\mathcal{O}\left(\rho\right)^{2},\\
g_{\phi\phi} & =&\ell^{2}(\xi^{+}+\xi^{-})^{2}+\ell^{2}(p^{+}+r^{-})(p^{-}+r^{+})
+\ell(p^{+}p^{-}-r^{+}r^{-})\rho+\frac{1}{2}(p^{+}p^{-}+r^{+}r^{-})\rho^{2}+\mathcal{O}\left(\rho\right)^{3}.
\end{eqnarray*}
This metric field is plugged in Einstein's equations
\begin{equation}\label{eeom}
R_{\mu \nu}-\frac{1}{2}R g_{\mu\nu}-\frac{1}{\ell^2}g_{\mu\nu}=0 \, ,
\end{equation}
leading to two copies of the AKNS integrable hierarchy, 
\begin{subequations}\label{akns-grav}
  \begin{align}\label{akns1}
   \pm \dot{r}^\pm+\frac{1}{\ell}\left(C'^\pm-2r^\pm A^\pm-2\xi^\pm C^\pm\right)&=0 \, , \\ \label{akns2}
    \pm \dot{p}^\pm+\frac{1}{\ell} \left( B'^\pm+2p^\pm A^\pm+2\xi^\pm B^\pm\right)&=0\, , \\ \label{akns3}
    A'^\pm-p^\pm C^\pm+r^\pm B^\pm&=0 \, .
  \end{align}
\end{subequations}

These solutions have two independent towers of AKNS conserved commuting charges \eqref{AKNS-Cargas}, associated with the ``left" and ``right" sectors. These charges could be understood as an example of ``soft hairs" at the horizon, in the sense of Hawking, Perry and Strominger \cite{Hawking:2016msc}, where the gauge transformations generated by $Q^\pm[\Lambda^\pm]$ satisfy an infinite
number of conservation laws, and also map a black hole configuration into physically inequivalent ones, without changing the energy of the system.

%Hence, the gauge transformations generated by $Q^\pm[\Lambda^\pm]$ map a black hole configuration into a physically inequivalent one, without changing the energy of the system.

\section{Flat connections, the first order problem and Wilson lines} \label{Section Flat}

In this section, we show how the associated first-order problem is connected with flat solutions of the zero curvature equation. We review properties of the Wilson lines, that are fundamental solutions of the linear problem and generate parallel transport. We study the evolution of the Monodromy matrix, which permits the construction of invariants through its trace.

\subsection{First order problem}
Let us consider a two-dimensional column vector $\Psi$ 
\begin{equation}
    \Psi=\begin{pmatrix}
\psi_{1}\\
\psi_{2}
\end{pmatrix}, 
\end{equation}
where $\Psi$ is a vector that stacks the solutions of the overdetermined linear system of equations,
\begin{equation}\label{linear eq}
    \partial_{t}\Psi=-V\Psi\quad\text{and}\quad\partial_{\varphi}\Psi=-U\Psi,
\end{equation}
being $U,V$ elements of the special linear group $SL(2,\mathbb{R})$.
The associated compatibility equation $\partial_{\varphi}\partial_{t}\psi=\partial_{t}\partial_{\varphi}\psi$ reduce to the zero curvature equation in two dimensions \eqref{dinamica asintotica}, where $V$ can be identified with the temporal component of a two-dimensional gauge field  $\mathcal{A}=\mathcal{A}_{\mu}dx^{\mu}$ and $U$ with its spatial component, such that $\mathcal{A}=Vdt+Ud\varphi$.
The zero curvature equation in 2D is immediately solved considering that the gauge field is flat, meaning that it can be recast as a linear equation of the form 
\begin{equation}\label{flat-eq}
 dg=-\mathcal{A}g.   
\end{equation}

If $\mathcal{A}$ transforms as a gauge connection $\mathcal{A}= b^{-1}\tilde{\mathcal{A}}b+b^{-1}db$, with $b\in \mathfrak{g}$, the first order problem \eqref{linear eq} is preserved when $\Psi$ satisfy the transformation law,
 \begin{equation} \label{transf G}
    \Psi= b^{-1} \tilde{\Psi}.
\end{equation}    
For the particular interest of this article, here $g$ and $b$ are also elements of  $SL(2,\mathbb{R})$.

\subsubsection{Hill's equation: Schr\"odinger equation for periodic potentials}
Let us consider the connection \eqref{eq:kdvbc}, associated to KdV boundary conditions. If we additionally define
\begin{equation}
 g=\begin{pmatrix}
\psi_{1} & \psi_{2}\\
-\psi_{1}' & -\psi_{2}'
\end{pmatrix},   
\end{equation}
the flat decomposition of the gauge field \eqref{flat-eq} along the spatial direction reduces to 
\begin{equation} \label{Schrodinger}
  \left(\partial^{2}_{\varphi}+\frac{2\pi}{k}\mathcal{L}\right) \psi_{1,2}=-\frac{\lambda^{2}}{2}\psi_{1,2}\, ,
\end{equation}
where $2\pi\mathcal{L}/k$ is the potential energy of the non-relativistic Schr\"odinger equation and $\psi_{1,2}$ are two linearly independent solutions, also sometimes called wave functions, nonetheless they are real and do not need to be square-integrable. We can identify the determinant of the group element $g$ with the Wronskian,
\begin{equation}
 W=\text{det}[g],   
\end{equation}
associated to the Hill's equation, where the condition of invertibility $\psi_{1}\psi_{2}'-\psi_{2}\psi_{1}'\neq 0$ is equivalent to require $\psi_{1}$ and $\psi_{2}$ to be two linearly independent solutions. Since we are dealing with matrix representations of the $\mathfrak{sl}(2,\mathbb{R})$ algebra, it is given that $W=1$. Then, the solutions $\psi_{1,2}$ are said to be “normalized".

Here $\lambda^{2}$ is the eigenvalue associated with the eigenfunction $\psi$, that in general, can be time-dependent. 

\textit{Proposition: }If $\cL$ is the potential of the KdV equation, then $\lambda$ is a constant parameter. 

\textit{Proof:} 
The proof stems by considering the Lax Pair structure of the KdV equation. We take the differential operators 
\begin{equation}
    L=\partial_{\varphi}^{2}+\frac{1}{6}u(t,\varphi)\qquad\text{and}\quad M=4\partial_{\varphi}^{3}+\frac{1}{2}(\partial_{\varphi}u(t,\varphi)+u(t,\varphi)\partial_{\varphi}),
\end{equation}
Evaluating them in the Lax evolution equation \eqref{laxeq}, one obtains the KdV equation
\begin{equation}
    \dot{u}=u\partial_{\varphi}u+\partial_{\varphi}^{3}u,
\end{equation}
 which can be scaled to have arbitrary coefficients in front of all the terms so as to recover the equation in the form presented previously \eqref{kdveq}. Having a Lax pair structure is enough to prove isospectrality, as shown in subsection \ref{subsection: Lax Pairs}. Then, $\lambda$ is constant in time.

The temporal component of \eqref{flat-eq} gives the time evolution for $\psi$, 
\begin{eqnarray}
    \dot{\psi}&=& -\psi(\lambda \epsilon+\frac{1}{2}\epsilon')-\psi'\epsilon\\
    \dot{\psi}'&=& \psi(-\frac{2 \pi  \epsilon \mathcal{L}}{k}+\lambda  \epsilon ' +\frac{1}{2} \epsilon'')-\psi'(\frac{1}{2} \epsilon '+\lambda  \epsilon)
\end{eqnarray}
The compatibility equation between the above set of equations and the spatial one \eqref{Schrodinger}, gives rise to the KdV-hierarchy equation \eqref{KdV hierarchy}.

\subsubsection*{Gravitational connections and the linear problem}

 When one takes the gauge fields $a$ to be related with asymptotic conditions in 2+1, as in \eqref{radial gaugetrans}, and one considers that the gravitational connection is also flat $dG=-AG$, the equation $G=b^{-1}g$ (being $b$ the radial dependent group element that gauges away the dependence on that coordinate) recovers the relation with the linear problem $dg=-ag$. 

For gravitational gauge connections \eqref{bdab}, we also have to consider that $dg^{\pm}=\mp a^{\pm}g^{\pm}$, in order to recover the Hill's equations \eqref{Schrodinger}. Then, the previous analysis is valid for the ``right" (+) sector, while for the (-) copy, the gauge group element has to be chosen as 
\begin{equation}
   g^{-}=\begin{pmatrix}
-\psi^{- \prime}_{1} & -\psi^{-\prime}_{2}\\
\psi^{-}_{1} & \psi^{-}_{2}
\end{pmatrix}.  
\end{equation}

\subsection{Wilson lines, monodromy and invariants}

\subsubsection{Wilson lines}
The system of equations \eqref{linear eq} is solved by 
\begin{equation}
    \Psi(t,\varphi)=\mathcal{P}\text{exp}\left(\int_{\gamma} \mathcal{A}_{\mu}dz^{\mu}\right)\Psi_{0},\qquad \Psi_{0}=\Psi(0,0)
\end{equation}
where $\mathcal{P}$ denotes a path-ordered operator acting on the matrix exponential, as the integrand is given by anticommuting elements when evaluated at different points in the curve. Here $\Psi_{0}=\Psi(0,0)$ is some initial condition, $\gamma$ is a curve in $\mathbb{R}^{1}\times S^{1}$, from $z_{i}=(t,\varphi)$ to $z_{f}=(0,0)$. The group element,
\begin{equation}
  \mathcal{W}(z_{i},z_{f})=\mathcal{P}\text{exp}\left(\int_{\gamma} \mathcal{A}_{\mu}dx^{\mu}\right)
\end{equation}
is the so-called \textit{Wilson line}, and represents the parallel transport of a vector $\Psi$  along a path $\gamma$ induced by the connection $A$, 
\begin{equation}
    \Psi_{\gamma}=\mathcal{W}\Psi.
\end{equation}
%The path ordering is explicitly given by
%\begin{align}
    %\mathcal{W}(z_{i},z_{f})=I+\int \frac{d z^{\mu}}{d}
%\end{align}
%after Taylor expanding the exponential 

It is also a two-point function, as it only depends on the initial and final points of the path. Another property of $\mathcal{W}$ is that it satisfies a superposition formula. Let us consider the path of integration as the sum of infinitesimal adjacent paths $\gamma=\gamma_{1}\cup\gamma_{2}\cup\cdots\cup\gamma_{n}$, where $\delta \gamma=\gamma_{i+j}-\gamma_{i}$, then
\begin{equation}
    \mathcal{W}_\gamma=\mathcal{W}_{\gamma_{1}}\mathcal{W}_{\gamma_{2}}\cdots\mathcal{W}_{\gamma_{n}}
\end{equation}
\textit{Proposition:} Under gauge transformations, the Wilson line satisfies the transformation law
 \begin{equation} \label{translaw wilson line}
   \mathcal{W}(z_{i},z_{f})\rightarrow b(z_{f}) \mathcal{W}(z_{i},z_{f}) b^{-1}(z_{i}).
\end{equation}   
\textit{Proof:} Let us consider 
\begin{equation}
  \Psi(z_{f})=\mathcal{P}\text{exp}\left(\int^{z_{i}}_{z_{i}} \mathcal{A}_{\mu}dz^{\mu}\right)\Psi(z_{i})
\end{equation}
We apply the transformation \eqref{transf G}, so that we have that $\Psi(z)\rightarrow b(z)^{-1}\tilde{\Psi}$. As a result
\begin{equation}
   b(z_{f})^{-1}\tilde{\Psi}(z_{f})=\mathcal{P}\text{exp}\left(\int_{\gamma} A_{\mu}dz^{\mu}\right) b(z_{i})^{-1}\tilde{\Psi}(z_{i}).
\end{equation}
Then
\begin{eqnarray}
    \tilde{\Psi}(z_{f})&=& b(z_{f})\mathcal{P}\text{exp}\left(\int_{\gamma} A_{\mu}dz^{\mu}\right) b(z_{i})^{-1}\tilde{\Psi}(z_{i})\\
     &=&\tilde{\mathcal{W}}(z_{f},z_{i})\tilde{\Psi}(z_{i})
\end{eqnarray}
where \eqref{translaw wilson line} is recovered.
\subsubsection{Monodromy matrix and conserved quantities}
  Let us consider 
  \begin{equation}
  \mathcal{W}(\lambda;z_{i},z_{f})=\mathcal{P}\text{exp}\left(\int_{\gamma} \mathcal{A}_{t}(\lambda;t,\varphi)dt+\mathcal{A}_{\varphi}(\lambda;t,\varphi)d\varphi\right)
\end{equation}
for $z=(t,\varphi)$ and $\lambda$ is the spectral parameter. As we have shown, if we consider $\mathcal{A}$ to be a flat connection, it will always satisfy the zero curvature condition, and the value of the path-ordered exponential is independent of the choice of the path. We choose $\gamma$ such that $\varphi\in [0,2\pi]$ with fixed time $t$. The matrix $M(\lambda)=\mathcal{W}(\lambda;0,2\pi)$
   \begin{equation}
  M(\lambda)=\mathcal{P}\text{exp}\left(\int_{0}^{2\pi} \mathcal{A}_{\varphi}(\lambda;\varphi)d\varphi\right)
\end{equation}
is called the \textit{Monodromy matrix}. Then, the Monodromy matrix is the fundamental solution of the periodic linear problem.

\textit{Proposition:} Given the monodromy matrix $M$, where
\begin{equation}
    H_{k}(\lambda)=\text{Tr}(M^{k}(\lambda))
\end{equation}
for $k$ a natural number, then all $H_{k}(\lambda)$ are independent of time, then generate conserved quantities.

\textit{Proof}: Let us compute the evolution equation of the Monodromy matrix, 
\begin{eqnarray*}
    \partial_{t}M(\lambda)&=&\int_{0}^{2\pi}\left[\mathcal{P}\text{exp}\left(\int_{\varphi}^{2\pi} \mathcal{A}_{\varphi}(\lambda;\varphi')d\varphi'\right)\right]\partial_{t}\mathcal{A}_{\varphi}(\lambda;\varphi)\left[\mathcal{P}\text{exp}\left(\int_{0}^{\varphi} \mathcal{A}_{\varphi}(\lambda;\varphi')d\varphi'\right)\right]d\varphi\\
    &=&\int_{0}^{2\pi}\left[\mathcal{P}\text{exp}\left(\int_{\varphi}^{2\pi} \mathcal{A}_{\varphi}(\lambda;\varphi)d\varphi'\right)\right]\partial_{\varphi}\mathcal{A}_{t}-[\mathcal{A}_{t},\mathcal{A}_{\varphi}]\left[\mathcal{P}\text{exp}\left(\int_{0}^{\varphi} \mathcal{A}_{\varphi}(\lambda;\varphi)d\varphi'\right)\right]d\varphi\\
    &=&\int_{0}^{2\pi}\partial_{\varphi}\left(\left[\mathcal{P}\text{exp}\left(\int_{\varphi}^{2\pi} \mathcal{A}_{\varphi '}(\lambda;\varphi)d\varphi\right)\right]A_{t}(\lambda;\varphi)\left[\mathcal{P}\text{exp}\left(\int_{0}^{\varphi} \mathcal{A}_{\varphi '}(\lambda;\varphi)d\varphi\right)\right]\right)d\varphi\\
      &=&A_{t}(\lambda;2\pi)M(\lambda)-M(\lambda)\mathcal{A}_{t}(\lambda;0)
\end{eqnarray*}
Considering periodic fields in the angle $\mathcal{A}_{t}(\lambda;\varphi)=\mathcal{A}_{t}(\lambda;\varphi+2\pi)$, 
\begin{equation}
    \partial_{t}M(\lambda)=[\mathcal{A}_{t}(\lambda;0),M(\lambda)].
\end{equation}
The Monodromy matrix satisfies the Lax pair equation. Provided the proof in subsection \ref{subsection: Lax Pairs},  $\text{Tr}(M(\lambda,t))$ is conserved in time. Indeed, when the Monodromy matrix is considered to be expanded in the spectral parameter $\lambda$,
\begin{equation}
    \text{Tr}(M(\lambda))=\sum_{i}Q_{i}\lambda^{i}\quad\text{with}\quad i \in\mathbb{N},
\end{equation}
 $Q_{i}$ are conserved charges. We provide now a concrete example.

\subsubsection{KDV Conserved quantities from the trace invariants}
We compute the conserved quantities of the KdV hierarchy, already obtained from the canonical charges of the Chern-Simons theory \eqref{canonical-kdv}, now from the invariants associated to the trace of the monodromy matrix. For simplicity, we look for a gauge transformation $g$ that diagonalizes the KdV boundary conditions \eqref{eq:kdvbc}, as it will simplify the computation. We consider the gauge connection  $\mathfrak{a}_{\varphi}=\cJ(\varphi) L_0$, where
\begin{equation} \label{gaugetransf2}
  \mathfrak{a}_{\varphi}=g^{-1} a_{\varphi} g+ g^{-1} \partial_{\varphi} g.
\end{equation}
In general, the Wilson line evaluated in two points $(\varphi,\varphi')$, transforms under gauge transformations as 
\begin{equation}
\mathcal{W}[\mathfrak{a}_{\varphi}]=g(\varphi)\mathcal{W}[a_{\varphi}]g^{-1}(\varphi')
\end{equation}
We demand the gauge transformation to be periodic in the angle, so that for a closed loop $g(0)=g(2\pi)$ and
\begin{equation}
{\rm tr}(\mathcal{W}[\mathfrak{a}_{\varphi}])={\rm tr}(\mathcal{W}[a_{\varphi}])
\end{equation}

We verify the existence of such gauge transformation \eqref{gaugetransf2}, that can be written as $g=e^{p\,L_0}e^{h\,L_1}e^{f\,L_{-1}}$, for some periodic functions $p(\varphi),h(\varphi),f(\varphi)$. Demanding only diagonal terms in the gauge transformation \eqref{gaugetransf2}, one get the equations
\begin{equation} 
  \cL=\frac{k}{2 \pi}\left(-2\lambda\,f+f^2-f'\right),\qquad f=\frac{1+2\lambda\,h-h'}{2h}.  
\end{equation}
These two conditions combined leads to $$\frac{8\pi}{k}\cL+4\lambda^2=\frac{1}{h^2}-\frac{h'^2}{h^2}+2\frac{h''}{h}.$$ 
In the diagonal part, one finds
\begin{eqnarray}\label{eqdeJ}
    \cJ=\frac{1-h'-h\,p'}{h}.
\end{eqnarray}
We relate the function $\cJ$ with the original field $\cL$, appearing in the gauge fields \eqref{eq:kdvbc} choosing $p=\frac{-2h'}{h}$, and noting that provided \eqref{eqdeJ},
\begin{equation} \label{expan J}
    \cJ^2+\cJ'=\frac{2\pi}{k}\cL+\lambda^2.
\end{equation}
This is a Miura transformation that can be understood as the Ricatti equation for $\cJ$. Then, the field on the diagonal gauge is related by a Miura transformation to the one defined in \eqref{eq:kdvbc}, that allows to construct a recursion relation that recovers the KdV Hamiltonian densities proposed in \eqref{canonical-kdv}. In general, this equation can be solved by considering the expansion,
\begin{equation} \label{expansionJ}
    \cJ=\lambda+\sum_{n=1}^{\infty} \cJ_{n}(2\lambda)^{-n},
\end{equation}
and solving order by order in $\lambda$. From this procedure, it is found that $\cJ_{0}=0$, $\cJ_{1}=2\pi\,\cL/k$, and 
\begin{equation} \label{Gerlfand2}
    \cJ_{n+1}=-\cJ_{n}'-\sum_{m=1}^{n-1}\cJ_{n-m}\cJ_{m},\quad n\geq1.
\end{equation}
The above is also an alternative recursion to generate the conserved quantities. Indeed, 

    \begin{equation}
    {\rm tr}(\mathcal{W}[\mathfrak{a}_{\varphi}])=2\cosh(H_{n})
\end{equation}
where $H_{n}$ are the KdV hamiltonians, such that
\begin{equation}
   H_{n}= \int \cJ_{2n+1} d\varphi\quad n\geq0.
\end{equation}
Terms of the recursion \eqref{Gerlfand2} with even subindices $\cJ_{2n}$ leads to total derivatives.

\section{Conclusions}

The idea of this manuscript was to present a unifying perspective that can be useful for understanding both asymptotic symmetries and properties of integrable systems. It was showed how integrable models in $D = 2$ also appear in 3D Chern-Simons theories, but under the notion of their asymptotic dynamics and asymptotic symmetries. We have classified the gauge transformations according to their action on the symmetry generators, which are the canonical charges of the theory. We compute the charges of the Korteweg-de Vries hierarchy and also those associated with the AKNS integrable system. We focus on the Korteweg-de Vries hierarchy, because of its deep connection with the Virasoro group, and recover the the KdV Hamiltonians from the invariants of the associated Monodromy matrix. In this respect, an important question remains, and that is the connection between these two points of view and how to prove the explicit relation between them, since they end up with the same results.

Along this lectures, we also propose new applications of integrability to General Relativity, providing a new arena of the exploration, apart from the Ernst equation \cite{ernst}, or the study of gravitational solitons \cite{gravsol}. It could help to reach various features of the theory using the tools available for integrable models, such as scattering problems \cite{Wiegmann,Abdalla,Bombardelli:2016scq} and other quantum features \cite{Retore:2021wwh}.

We consider that bringing non-linear asymptotic dynamics to gauge theories, in a way that is controlled by the inherent infinite symmetries of integrability, could help to introduce interesting new phenomena (e.g. non-linear Hall effects, related with integrability and Chern-Simons \cite{Weigmann2}, or with geometric phases \cite{Oblak:2020bkg}), even more so if it is governed by solvable models with great application to fluid dynamics and condensed matter.

\section{Acknowledgments}
It is a pleasure to thank the organizers of the school, Clara Aldana,
Camilo Arias Abad,
Diego Gallego Mahecha,
Carolina Neira Jimenez and
Sylvie Paycha for their support and fantastic work. The author also thanks to ``Universidad Nacional de Colombia" and the ``Centre International de Mathématiques Pures et Appliquée" for financial support. 
Special thanks to H. A. Gonz\'alez, for the careful reading  this manuscript.

%%%%%%%%%%%%%%%%%%%%%%%%%%%% BIBLIOGRAPHY %%%%%%%%%%%%%%%%%%%%%%%%%%%%%%%%%%%%%%

%%%%%%%%%%%%%%%%%%%%%%%%%%%%%%%%%%%%%%%%%%%%%%%%%%%%%%%%%%%%%%%%%%%%%%%%%%%%%%

\end{document}